# Atomic layer deposition and superconducting properties of NbSi films

*Thomas Proslier*, Jeffrey A. Klug, Jeffrey W. Elam,*

*Helmut Claus, Nicholas G. Becker, Michael J. Pellin*

Argonne National Laboratory, 9700 S. Cass Avenue, Argonne IL 60439.

prolier@anl.

Atomic layer deposition was used to synthesize niobium silicide (NbSi) films with a 1:1 stoichiometry, using $NbF_5$ and $Si_2H_6$ as precursors. The growth mechanism at 200°C was examined by *in-situ* quartz crystal microbalance (QCM) and quadrupole mass spectrometer (QMS). This study revealed a self-limiting reaction with a growth rate of 4.5 Å/cycle. NbSi was found to grow only on oxide-free films prepared using halogenated precursors. The electronic properties, growth rate, chemical composition, and structure of the films were studied over the deposition temperature range 150-400°C. For all temperatures, the films are found to be stoichiometric NbSi (1:1) with no detectable fluorine impurities, amorphous with a density of 6.65g/cm$^3$, and metallic with a resistivity ρ=150 μΩ.cm at 300K for films thicker than 35 nm. The growth rate was nearly constant for deposition temperatures between 150-275°C, but increases above 300°C suggesting the onset of non-self limiting growth. The electronic properties of the films were measured down to 1.2K and revealed a superconducting transition at $T_c$=3.1K. To our knowledge, a superconducting niobium silicide film with a 1:1 stoichiometry has never been grown before by any technique.

## Introduction:

Niobium silicide based alloys have a wide range of applications including tunnel barriers for Josephson junctions [1], superconductors for particle detection (bolometers) [2-3], and low friction and high temperature corrosion resistant coatings for engines [4]. Thin films of $Nb_xSi_{1-x}$ compounds with various compositions can be deposited using a variety of methods including electron beam evaporation [5-6] ($Nb_5Si_3$ and $NbSi_2$), RF magnetron sputtering [7], explosive or arc melting and chill-casting [8-10] ($NbSi_3$), chemical vapor deposition [11], direct laser fabrication [12], or ion-induced formation [13]. In contrast with these line of sight techniques, atomic layer deposition (ALD) offers many advantages. ALD is a thin film deposition technique that uses alternating, self-limiting chemical reactions between gaseous precursors and a solid surface to deposit materials in an atomic layer-by-layer fashion [14]. The self-terminating chemistry coupled with gaseous diffusion of the precursor vapors allows an excellent control of both the composition and the thickness uniformity on arbitrary complex-shaped objects. ALD has been used previously to deposit a wide range of materials including metals, metal oxides, metal nitrides but only metal silicides have received relatively little attention [15]. To our knowledge this is the first reported in-depth growth study for the ALD of a pure silicide.

There has been only one previous report for the ALD of superconducting materials [16]. Despite the fact that very little effort has been invested in the growth study of superconductors by ALD, this unique technique could clearly beneficiate a variety of low temperature superconductor-based applications that are currently limited by line-of-sight deposition techniques. Bolometers are one example [2-3]. Another example is superconducting radio frequency (SRF) accelerating niobium cavities [17] used is almost all particle accelerators around the world.

This study pursues dual goals; first to develop a low temperature, reliable ALD process to synthesize NbSi with a high growth rate that can open the way to other transition metal silicide or NbSi-based alloys growth study by ALD and second to characterize and optimize the superconducting properties of the NbSi films. In this study we use alternating exposures to $NbF_5$ and $Si_2H_6$ to deposit NbSi films by ALD. In situ quartz crystal microbalance (QCM) and quadrupole mass spectrometer (QMS) measurements are used to explore the mechanism for the NbSi ALD. A variety of ex-situ techniques is used to determine the physical properties of the films. X-ray reflectivity (XRR) measurements are used to evaluate the thickness, roughness, and density of the films while Rutherford Backscattering Spectroscopy (RBS) is employed to determine the density and stoichiometry of the films. The composition and impurity content of the films is evaluated using X-ray photoelectron spectroscopy (XPS) and scanning electron microscopy (SEM) is utilized to examine the uniformity and conformality of the coatings on non-planar surfaces. The crystallinity of the films was examined using X-ray diffraction (XRD). The electrical properties were measured using four-point probe resistance and SQUID magnetometry down to 1.2 K.

**Experimental details:**

The ALD experiments were conducted using a custom viscous flow reactor in which the deposition zone is an Inconel 600 tube of 2" diameter. The reactor temperature was controlled by an external resistive 3-zone system and measured in 9 spots along the deposition tube to insure homogeneity. Ultra pure Nitrogen (99.999%) further purified by an oxygen filter (Aeronex Gatekeeper) was used as a carrier and purge gas. The total $N_2$ flow was maintained at 360 sccm using mass flow controllers and the average pressure inside the ALD apparatus was ~ 1.3 Torr. The reactor was also equipped with a differentially pumped quadrupole mass spectrometer (QMS) (Stanford research Systems, Model RGA300) located downstream from the sample location and separated from the reactor by a 35-μm pin-hole. A quartz crystal microbalance (QCM) was used to monitor the ALD growth *in situ*. The QCM used a Maxtek Model BSH-150 sensor head accommodating a single side polished quartz crystal sensor (Tangidyne/VB) and interfaced to the computer via a Maxtek Model TM400 film thickness monitor.

The NbSi films were deposited on quartz, silicon (100) or sapphire substrates using alternating exposures to niobium (V) fluoride ($NbF_5$, 98%, Sigma-Aldrich) and disilane ($Si_2H_6$, 99.998% Sigma-Adrich). Some of the experiments utilized silane ($SiH_4$, 99.998%, Sigma-Aldrich) in place of the $Si_2H_6$. The $NbF_5$ was held at 65°C in a stainless steel bubbler while the $Si_2H_6$ and $SiH_4$ compressed gases were contained in stainless steel lecture bottles. To quantify our QMS signals and also to nucleate the NbSi growth, we performed in-situ measurements of W ALD using alternating exposures to tungsten hexafluoride ($WF_6$, 99.9%, Sigma-Aldrich) and $Si_2H_6$. We also used alternating exposures to trimethyl aluminum (TMA, Sigma-Aldrich, 97%) and deionized water to prepare ALD $Al_2O_3$ films as a nucleation layer for the ALD W.

Each NbSi ALD cycle consisted of a $NbF_5$ exposure for time t1, a $N_2$ purge for time t2, a $Si_2H_6$ exposure for time t3, and a second $N_2$ purge for time t4. The timing sequence is expressed as (t1-t2-t3-t4) where all times are in seconds. The partial pressures for the $Si_2H_6$ and $NbF_5$ during the exposure times for these precursors were 0.15 and 0.02 Torr, respectively, as measured using a Baratron capacitance manometer.

The chemical composition was measured ex-situ by X-ray photoemission spectroscopy (XPS, Perkin Elmer Φ500) and Rutherford back scattering (RBS, Evans Analytical). The film thickness and density was determined by X-ray reflectivity (Expert-Pro MRD, Philips, Cu Kα x-rays ) and the structural analysis by X-ray diffraction (Rigaku Model ATXG rotating anode, Cu Kα x-rays). Scanning electron microscope (SEM) images were conducted on a Hitachi Model S4700 system. The electronic and superconducting film properties were measured by a four point probe method at room temperature (model Lucas Signatone S-301-6 probe stand with a Keithley 224 current source) and with a home-made superconducting quantum interference device (SQUID) down to 1.2K under an external field of 10 mGauss respectively.

**Results and discussion:**

***In-situ* studies.**

We performed *in-situ* QCM and QMS measurements at 200 $^oC$ using the timing sequence: 2-10-1-10 to explore the mechanism for NbSi ALD using $NbF_5$ and $Si_2H_6$. The QMS measurements provide the identities and relative quantities of the gaseous reaction products released during the individual half-reactions of each ALD cycle. The QCM measurements yield the growth rate and the relative mass of the surface species remaining after each half reaction. When combined with ex-situ measurements of the NbSi composition, the QCM and QCM results give an accurate description of the ALD surface chemistry during each half-reaction and provide insight into the mechanism for film growth.

Representative QMS data recorded during the $NbF_5$ and $Si_2H_6$ exposures are shown in Fig. 1. This plot shows two complete ALD cycles followed by 4 exposures to only $Si_2H_6$ and then 4 exposures to only $NbF_5$. The exposures to just a single precursor were utilized to evaluate the background signals for each of the mass peaks so that these can be subtracted to yield the signals arising solely from the gaseous products of the ALD surface reactions.

A peak at mass-to-charge ratio, *m/z* = 104 appears during the $NbF_5$ half reactions but not during the $Si_2H_6$ half-reactions. However, when no $Si_2H_6$ is pulsed, the *m/z* = 104 peak associated with the $NbF_5$ disappears as shown by the final five ALD cycles in figure 1. From these observations we conclude that the peak at *m/z* =104 represents a product of the $NbF_5$ ALD half-reaction. By collecting similar QMS data and background signals over the mass range 2-110 amu, we discovered that the $NbF_5$ reaction yields the following products (and relative abundances) at *m/z* = 104 (1.5), 87 (5.5), 86 (12), 85 (100), 47 (10), and 33 (7). This mass pattern matches closely the fragmentation pattern for silicon tetrafluoride ($SiF_4$) [18]. A similar trend is found for *m/z* = 20 that corresponds to hydrogen fluoride (HF). From these observations we conclude that $SiF_4$ and HF are the only gas phase products during the $NbF_5$ half reaction.

During the $Si_2H_6$ exposures for NbSi ALD we see a sharp spike at mass *m/z* = 2 coincident with the leading edge of the $Si_2H_6$ pulses followed by a smaller plateau that persists as long as the $Si_2H_6$ valve is kept open as illustrated by the first 3 ALD cycles in Fig. 1. However, when only $Si_2H_6$ is pulsed, the sharp spike is absent as shown by ALD cycles 4-6 in Fig. 1. From this we conclude that the sharp spike represents $H_2$ liberated as a gas-phase product of the $Si_2H_6$ half-reaction. In addition, a series of peaks appears at *m/z* = 47(30), 66(15), 67(65.8), 85(100), and 86 (10) during the $Si_2H_6$ exposures, but no companion peak is at *m/z* = 104. These peaks match the cracking pattern of trifluorosilane ($SiHF_3$) [19], and we observed that they also disappear when $Si_2H_6$ is pulsed repeatedly. We can therefore conclude that $H_2$ and $SiHF_3$ are the only gaseous reaction products of the $Si_2H_6$ half-reaction.

In order to quantify the ALD NbSi QMS signals, we performed in-situ measurements during W ALD. The W ALD process is well-characterized and the mechanism is understood [20, 21, 22]. Consequently, the QMS signals observed during W ALD can be used to calibrate the corresponding QMS signal observed during NbSi ALD. Using alternating exposures of $WF_6$ and $Si_2H_6$ at 200 $^oC$, W deposits at 1000 ng/cm$^2$ per cycle which equates to $5.3 \times 10^{-9}$ mol W/cm$^2$ [23]. We measured a growth rate of 300 ng/cm$^2$ per cycle by QCM during NbSi ALD (Fig. 2) which equates to a concentration of only $2.5 \times 10^{-9}$ mol Nb/cm$^2$. During W ALD, one $SiF_3H$ molecule is released per adsorbing $Si_2H_6$ molecule during the $Si_2H_6$ exposures [20]. We observed that the *m/z*=85 signal during the $Si_2H_6$ exposures for NbSi ALD were ~1/2 the intensity of the corresponding signals during W ALD so that ~ one $SiF_3H$ molecule is released per $Si_2H_6$ molecule during the $Si_2H_6$ half-reactions for NbSi ALD. Next we compared the peak heights observed during NbSi ALD for $SiF_3H$ at *m/z*=85 during the $Si_2H_6$ half-reaction with that of the $SiF_4$ peak at *m/z*=85 during the $NbF_5$ half-reaction. After correcting for the relative sensitivity factors we concluded that these products were formed in the ratio $SiF_4$:$SiF_3H$ ~ 1.2:1. Finally, we applied a similar analysis to conclude that the ratio for $H_2$:$SiF_3H$ ~ 6:1.

The composition of the films in this study was NbSi = 1:1 as revealed by ex-situ RBS measurements (see below). Combining this finding with the gas phase products and relative product ratios described above, and applying principles of mass balance, we postulate the following mechanism for NbSi ALD:

$$NbF_3^* + 11/6\ Si_2H_6 \rightarrow NbSi_3H_4F^* + 2/3\ SiHF_3 + 19/6\ H_2 \quad (1)$$

$$NbSi_3H_4F^* + 2\ NbF_5 \rightarrow (NbSi)_2NbF_3^* + SiF_4 + 4\ HF \quad (2)$$

In these equations the asterisks designate the surface species, and we have assumed that the NbSi surface is fluorine-terminated following the $NbF_5$ exposures in analogy with W ALD [24, 25]. The intermediate surface species $NbSi_3H_4F^*$ implies either multiple Nb-Si or Si-Si bonds. The true stoichiometry of this intermediate surface species could be evaluated using in situ infrared absorption measurements [22].

To check the validity of the reaction mechanism given by Eqns. 1 and 2, we analyzed the in situ QCM data recorded simultaneously with the QMS measurements. Figure 2a demonstrates that alternating $NbF_5/Si_2H_6$ exposures resulted in a linear mass increase versus time at a rate of 300 ng/cm$^2$/cycle. Figure 2b is an expanded view of the QCM data for 2 ALD cycles. During the $Si_2H_6$ exposures we saw a net mass increase $\Delta m_1$, = 50 ng/cm$^2$. There was an abrupt mass increase during the $NbF_5$ exposures followed by a slower apparent decrease in mass over the 10 s purge to a steady state value of $\Delta m_2$ = 300 ng/cm$^2$. We attribute the apparent mass loss during the $NbF_5$ purge to a slow temperature relaxation of the QCM surface back to thermal equilibrium following the transient heating induced by the exothermic reaction of the $NbF_5$. The QCM data yields a mass ratio $R = \Delta m_2/\Delta m_1 = 5.3$ averaged over 50 cycles. From Eqns. 1 and 2 and the known atomic weights, we predict $R$=5 in good agreement with the experimental value. This agreement lends support to the NbSi ALD mechanism postulated Eqns. 1 and 2. It is worth mentioning that we also succeeded in depositing NbSi films using silane ($SiH_4$) in place of $Si_2H_6$ as the reducing agent. QCM measurements revealed a step ratio $R$ = 5 and a growth rate of 240 ng/cm$^2$/cycle while X-ray reflectivity showed a density of 6.65 g/cm$^3$. These results are very similar to the values obtained using $Si_2H_6$ suggesting a similar reaction mechanism. A detailed study of NbSi ALD using $SiH_4$ is underway.

It is surprising that different metal fluoride precursors used in conjunction with disilane or silane can sometimes produce silicides (e.g. $NbF_5$, $TaF_5$) or other times pure metals (e.g. $WF_6$ or $MoF_6$). To investigate this phenomenon, we calculated the free energy changes for converting metal fluorides $NbF_5$, $TaF_5$, $WF_6$ and $MoF_6$ to either the corresponding silicides or the pure metal using a commercial software package [26]. Figure 3 reveals that the silicides are thermodynamically favored over the pure metals in all cases. However, whereas the difference in free energy between the metal and the silicide is small for W and Mo (24 and 13 kcal/mol respectively at 250 $^{o}$C), this difference is much larger for the Ta and Nb (both ~ 86 kcal/mol at 250 $^{o}$C). In other words, even though silicides formation is favorable in all cases, the thermodynamic driving force is much greater for $NbF_5$ and $TaF_5$, the precursors which are found experimentally to form silicides in ALD, as compared to $WF_6$ or $MoF_6$ that have been found to produce metal films. Although this simple analysis ignores many of the complexities of true ALD processes such as the formation of metastable products due to kinetic constraints, the trend in Fig.3 are very clear and are likely the primary reason for NbSi formation in our experiments as opposed to pure Nb.

**Nucleation and Growth of NbSi films.**

When we first attempted to deposit ALD NbSi films on Si(100), fused silica, and sapphire substrates at temperatures below 200°C, XRR analysis and visual inspection revealed that films did not grow on the bare substrates. We also attempted to deposit the NbSi films on substrates coated with ALD $Al_2O_3$ and $Nb_2O_5$ as well as sputtered Fe, Pt, NbTiN, and Nb that had been exposed to air, but in all cases the films did not grow. However, we did succeed in growing silver-colored NbSi films on metallic surfaces that had not been exposed to air. These metallic surfaces were prepared in our ALD reactor and included W films deposited using $WF_6/Si_2H_6$ as well as NbC and NbCN films prepared using $NbF_5$ and various reducing agents ($Si_2H_6$, $SiF_4$, TMA, and $NH_3$) [27]. From these observations we hypothesize that the NbSi growth is inhibited at lower temperatures on oxide surfaces such as the ALD metal oxides and the native oxides of the sputtered metal films formed by air exposure. At deposition temperatures above 200$^{o}$C the NbSi films could be deposited on any substrate following a short incubation time that depended on temperature. For instance, at 225$^{o}$C the nucleation delay was 5 cycles and above 250$^{o}$C no nucleation delay was seen. We hypothesize that the NbSi growth on oxide surfaces at these higher temperatures was initiated by the thermal decomposition of $Si_2H_6$ [28].

We performed additional QCM measurements to understand the inhibited NbSi growth at 200°C. The QCM measurements demonstrated that alternating exposures of $NbF_5$ and either $H_2O$ or $H_2O_2$ permitted niobium oxide ALD at growth rates of 2.0 and 2.3 Å/cycle, respectively. Upon substituting the $H_2O$ with $Si_2H_6$, the film growth stopped abruptly after just one $Si_2H_6$ exposure following the $NbF_5$ pulse. No growth was observed during the next ~100 cycles of alternating $NbF_5$ and $Si_2H_6$ exposures. However, when the $Si_2H_6$ was switched back to $H_2O$, the $Nb_2O_5$ growth resumed after just 2-3 cycles of $NbF_5$ and $H_2O$. We postulate that the Si etching necessary for the NbSi ALD (Eqn. 2) is inhibited by the formation of a stable $Nb_xSi_yO_z$ oxide.

For the remainder of our studies we deposited the ALD NbSi films after first performing 30 cycles of $Al_2O_3$ ALD followed by 15 cycles of W ALD to prepare a fresh metallic surface on the substrate. Figure 4 shows the results of uptake measurements made at 200°C while varying the $NbF_5$ exposure time using the timing sequence x-5-1-5. This figure demonstrates the self-limiting behavior for the $NbF_5$ reaction with saturation coverage achieved after exposure times of ~1s and a growth rate of 4.5 Å/cycle. The delay time of ~0.5 s preceding measurable NbSi growth in Fig. 4a results from the complete consumption of the $NbF_5$ precursor on the walls of the flow tube upstream of the substrate in our travelling wave reactor. Figure 4b shows a similar graph exploring the effect of increasing $Si_2H_6$ exposures using the timing sequence 2-5-x-5 and reveals nearly complete saturation after $Si_2H_6$ exposure times of only 0.25 s. The NbSi growth rate continues to increase somewhat for longer exposure times at a rate of ~ 0.15Å/s indicative of a small non-self-limiting component to the growth. This behavior probably results from the thermal decomposition of $Si_2H_6$ as has been noted previously for W ALD using $WF_6$ and $Si_2H_6$ [29]. The timing sequence 2-5-1-5 was used for the remainder of our studies and produced films with very uniform properties across multiple substrates placed along the 20 cm deposition zone in our ALD reactor.

The effect of deposition temperature on the NbSi growth rate and film roughness as determined from XRR measurements of films prepared using 100 NbSi ALD cycles on Si(100) substrates is shown in figure 5a. No growth was observed for a deposition temperature of 100°C, whereas the growth rate was constant at 4.5 Å/cycle over the temperature range 150 to 275°C. The growth rate drops slightly at 300°C and then increases substantially at higher temperature. Simultaneously the NbSi growth becomes non-uniform along the deposition zone of the ALD reactor. We attribute these behaviors to an increased component of non-self-limiting CVD at deposition temperatures above 275°C. The roughness increased dramatically from 0.4 nm at 150°C to 1.7 at 275°C as shown in Fig. 5a. Increases in surface roughness with growth temperature can result from CVD or from an amorphous to crystalline transition where the roughness results from preferred growth along certain crystalline planes. However, the NbSi film thickness is constant over much of this temperature range and the films appeared amorphous by X-ray diffraction (XRD). The increased roughness may result from an increasing number of NbSi crystals at higher deposition temperatures which are too small to be detected by XRD.

**Composition and Morphology of NbSi films.**

Both XRR and RBS measurements showed that the density of the NbSi films remained nearly constant in the range 6.5-6.7 over the full range of deposition temperatures as shown in Fig. 5b. The NbSi growth rate of 4.5 Å/cycle and density of 6.65 g/cm$^3$ at 200°C measured by ex situ XRR match very closely the growth rates determined using in situ QCM. Moreover, the film thicknesses and properties were found to be homogeneous upstream and downstream of the QCM position in the ALD reactor. These findings lend confidence to the conclusions drawn from our QCM measurements.

RBS measurements were also used to evaluate the composition of the NbSi films. Figure 6a shows that for all temperature studied the chemical composition of the films had a ratio of Nb:Si =1:1 over the ~1μm$^2$ RBS sampling zone. This finding is somewhat surprising since the XRR measurements revealed CVD at deposition temperatures above ~275°C. The consistent composition even in the CVD regime suggests that a similar growth mechanism consisting of Si deposition during the $Si_2H_6$ exposures and partial Si etching during the $NbF_5$ exposures (Eqns. 1 and 2) operates in both the CVD and ALD regimes.

*Ex-situ* XPS performed on the NbSi films without Ar sputtering (dot-dashed trace in Fig. 6b) revealed the presence of native silicon oxides (e.g. $SiO_2$). However, almost no niobium oxides were observed before Ar sputtering (dot-dashed trace of Fig. 6c) as demonstrated by the absence of features in the 207.8-210.5 eV region of the Nb 3d spectrum (region bounded by dashed lines in Fig. 6c). This behavior was observed over the full range of deposition temperatures and even when the films were exposed to room air for one week. This observation is surprising because usually niobium oxides form immediately upon air exposures. The final ALD precursor exposure was $Si_2H_6$ for all of the films in this series. Therefore, it is possible that a silicon-based passivating layer protected the underlying NbSi from oxidation. In support to this explanation, one sample prepared at 250 $^0$C used $NbF_5$ as the final precursor exposure, and in this case XPS revealed $Nb_2O_5$ on the surface that disappeared after 2 min of Ar sputtering.

After a mild Ar sputtering, the XPS spectrum showed oxide- and fluorine-free NbSi films (solid traces in Figs 6a and 6b). The films were also fluorine-free following Ar sputtering as evidenced by the lack of features in the F 1s region of the XPS survey scan (not shown.) The binding energy for the Nb 3d peak is 202.9 eV and is close to the binding energy of 202.4 eV expected for elemental Nb. The binding energy of the 1s Si peak is 99.2 eV and is also close to the expected binding energy for elemental Si of 98.6 eV. The binding energy for both the Nb 3d and Si 1s peaks are in agreement with previous XPS measurements of Nb silicides [30] and demonstrate that the Nb and silicon atoms are chemically bound together. The Nb and Si binding energies remain nearly constant as a function of the growth temperature (Fig. 6c), which further demonstrates that the composition and chemical binding remain constant with deposition temperature.

XRD data recorded from NbSi films deposited on fused silica substrates were featureless for all deposition temperatures implying an amorphous structure. However, SEM images of the top surface of the NbSi films prepared on Si(100) surfaces at 200°C using 300 NbSi cycles (Fig. 7b) showed small bumps consistent with tiny nanocrystals. These features may explain the surface roughness detected by XRR. Cross-sectional SEM images of NbSi films deposited on silicon trench structures under similar conditions (Fig. 7a) revealed that the films are highly uniform, conformal, and pinhole-free.

**Electrical Properties of NbSi films.**

Figure 5b shows that the NbSi resistivity is nearly constant at 150-190 μΩ.cm as a function of the growth temperature, as expected in light of the constant film properties detailed earlier. The temperature dependence of the resistance between 2-50K measured on a NbSi film grown on a Si(100) substrate using 100 ALD cycles at 200$^{o}$C is shown in Fig. 8. The graph shows an abrupt decrease below the critical temperature $T_c$=3.1K characteristic of a superconducting transition. To our knowledge, this is the first example of a superconducting niobium silicide film with a 1:1 stoichiometry.

SQUID determination of $T_c$ is more informative than the resistivity measurement since the width of the transition from the normal to the superconducting state indicates the homogeneity of the superconducting properties throughout the sample. A very sharp transition means a unique $T_c$ and therefore a unique alloy stoichiometry. In contrast, resistivity only measures the current flowing between 2 or 4 contacts and therefore the percolation path with the highest $T_c$, making the measurement insensitive to other phases. Figure 9 represents the sample magnetization M, as measured by SQUID, under an external magnetic field of 10 mGauss applied perpendicular to the sample as a function of the temperature.

Figure 9a shows SQUID measurements for the multilayer structure: Si(100)/$Al_2O_3$(30 cycles)/W(15 cycles)/NbSi(100 cycles)/$Al_2O_3$(21 cycles) grown at 200$^{o}$C. The blue curve labeled "as-grown" reveals a broad superconducting transition below 3.1K (indicated by the horizontal dashed line), which corresponds to the maximal $T_c$ of this film and coincides with the $T_c$ measured by resistivity. In comparison, the measurement done on a NbSi films grown at 400$^{o}$C in the CVD regime on 31 cycles of ALD $Al_2O_3$ (Fig. 9a, black dot-dashed curve) is much sharper. Similar results were obtained on NbSi films grown at 225 $^{o}$C and 275$^{o}$C as shown by the blue traces in Figs. 9b and 9c, respectively.

The broad distribution of $T_c$ values for the NbSi films grown in the ALD regime are consistent with a distribution of film compositions. Although RBS found a uniform, 1:1 NbSi composition at various locations on the samples, we can't exclude the possibility of compositional inhomogeneities that are smaller than the ~1 μm$^2$ RBS probe size. In addition, hydrogen trapped inside the films is known to have detrimental effects on the superconducting properties [31-35] and could also be the cause for the transition broadening. In this case, post-deposition annealing in inert gas could reduce the transition width by causing the hydrogen to diffuse out of the film. To test this idea, we annealed these samples into an Ar or $N_2$ atmosphere for 5 hrs. at 400$^o$C and measured again the superconducting transition by SQUID. The green curves in Figs. 9a-c (dashed green curve is for the 400$^o$C $N_2$ anneal, solid green curve is for the 400$^o$C Ar anneal) show much sharper transitions following the 400°C treatments indicating more homogeneous superconducting properties in the films. The chemical composition, measured by XPS, showed neither contamination nor diffusion into the NbSi layers following the 400°C treatments. These findings argue that H was indeed responsible for the broad transitions in the as-deposited NbSi films.

The NbSi films grown at 200$^o$C and capped in-situ with ALD $Al_2O_3$ were annealed to the higher temperature of 600 $^o$C for 5 hrs in Ar or $N_2$ atmosphere (Fig. 9a:the red solid curves is for 600$^o$C in Ar and the red dashed curve is for 600$^o$C in $N_2$). Following this treatment, XPS revealed the presence of oxygen in the NbSi. Furthermore, SQUID measurements found that the superconducting transition was severely decreased down to 2K (dashed red line in Fig. 9a). The color of this film remained silverish. For uncapped films grown at 225 and 275$^o$C the $T_c$ is suppressed after annealing at 600 $^o$C, and the films are found to be insulating. We also noticed a change in color from metallic to purple. A possible explanation for these changes is the incorporation of nitrogen in the uncapped NbSi films during the annealing at 600 $^o$C to form an insulating silicon/Niobium nitride alloy. Further XPS analysis would be needed to evaluate this possibility.

**Conclusions:**

We have presented a method for preparing NbSi films by ALD using alternating exposures to $NbF_5$ and $Si_2H_6$. The NbSi growth follows a mechanism similar to that of W in which Si deposited during the $Si_2H_6$ exposure is etched and replaced by Nb during the subsequent $NbF_5$ exposure. The gaseous byproducts were found to be $SiF_3H$ and $H_2$ during the $Si_2H_6$ half-reaction and $SiF_4$ and HF during the $NbF_5$ half-reaction. This method yields a growth rate of 4.5 Å/cy over a temperature range 150—300$^o$C.

The films were found to be pure NbSi with a 1:1 ratio and were metallic and superconducting below 3.1K over the growth temperature range 150-400°C. To our knowledge, this is the first example of a superconducting niobium silicide film with a 1:1 stoichiometry. Finally, we successfully improved the transition width and the homogeneity of the superconducting properties using a post-deposition annealing treatment in Ar or $N_2$ at 400°C for 5 hrs. The growth of thin and homogeneous superconducting films on arbitrary, complex-shaped objects could enable the development of 3-D bolometers or other superconductor-based devices. Further work on mixed-metal alloys combining the chemistries for W and NbSi ALD is ongoing, and we are also exploring the differences between the $SiH_4$ and $Si_2H_6$ reducing agents for the NbSi ALD.

**Acknowledgments:** The work was supported by the U.S. Department of Energy, Office of Science under contract No. DE-AC02-06CH11357 and by the American Recovery and Reinvestment Act (ARRA) through the US Department of Energy, Office of High Energy Physics Department of Science. The electron microscopy was accomplished at the Electron Microscopy Center for Materials Research at Argonne National Laboratory, a U.S. Department of Energy Office of Science Laboratory operated under Contract No. DE-AC02-06CH11357.

**Figures**

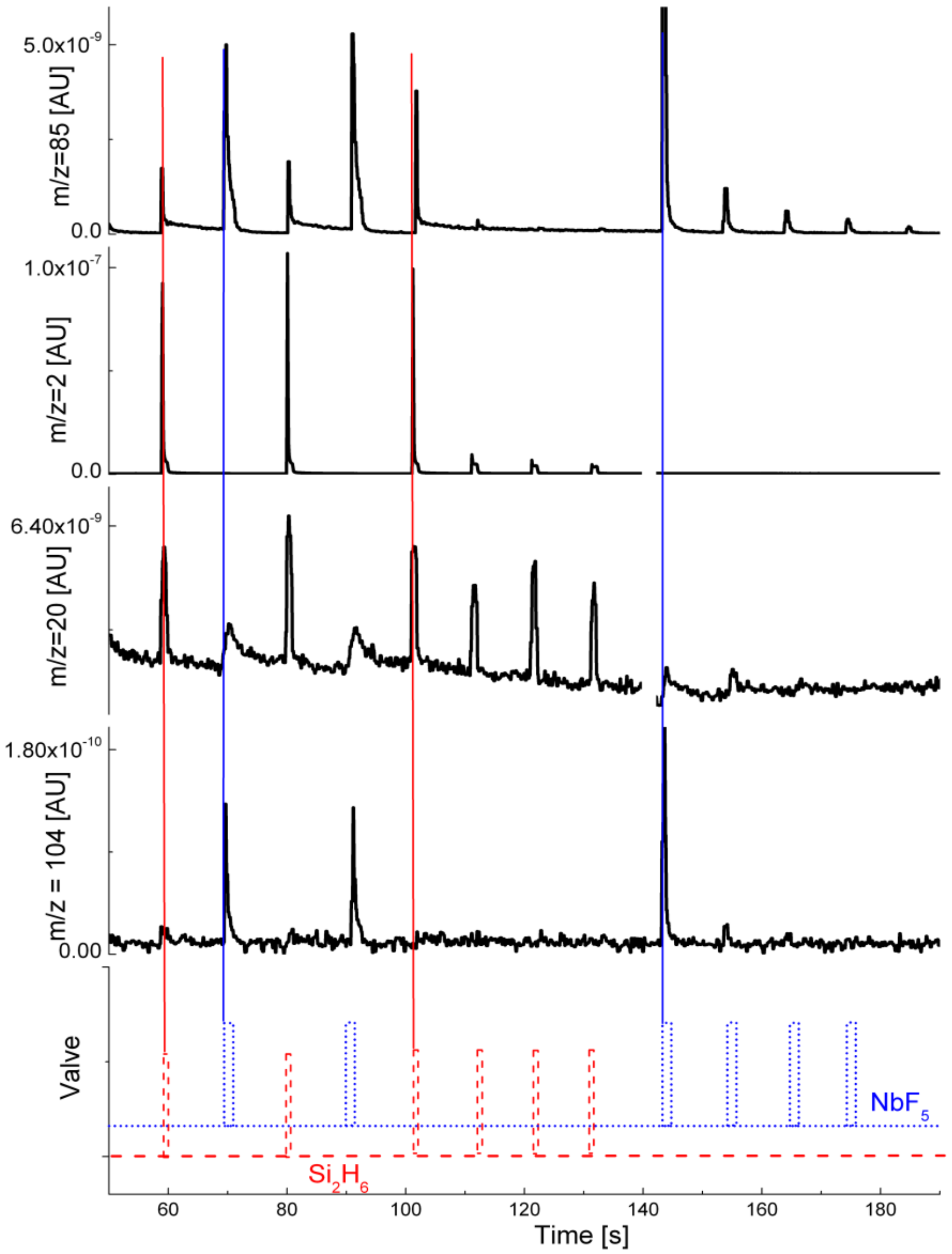

Figure 1: From top to bottom, QMS signal for *m*=85 to 104 versus time during alternating exposure to $NbF_5$ (dotted line) and $Si_2H_6$ (dashed line) at 200$^o$C using the timing sequence 2-10-1-10.

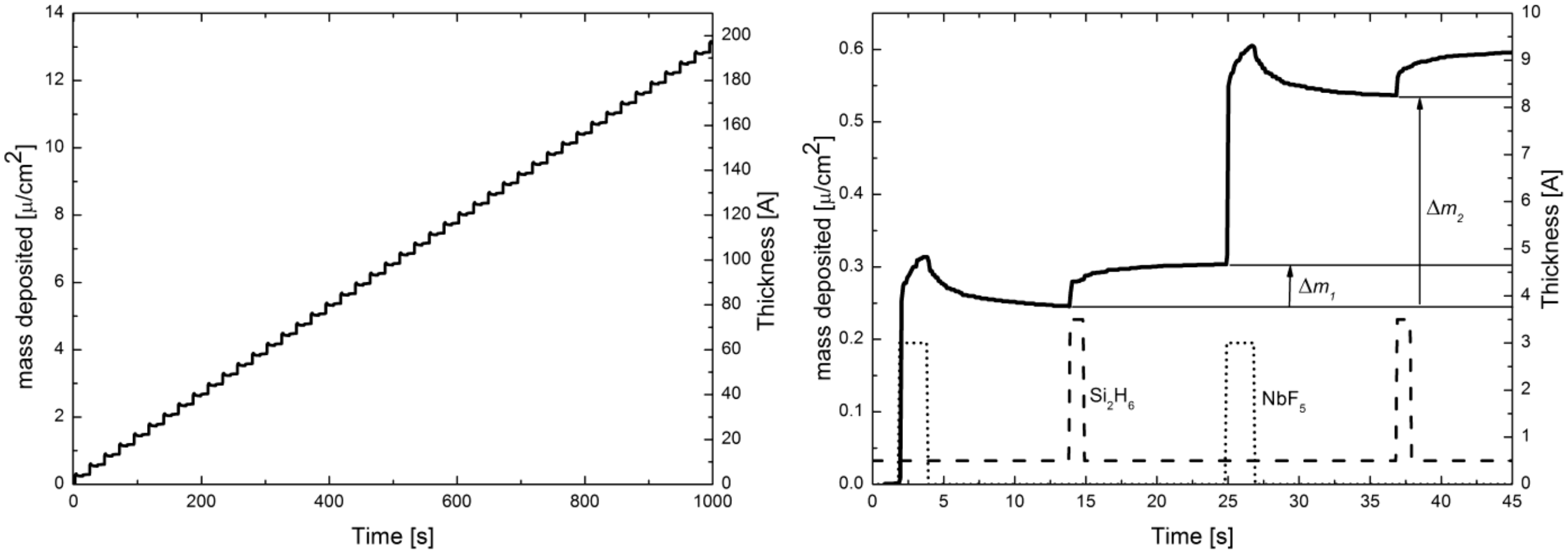

Figure 2: (a) QCM signal versus time during the alternating exposures to exposure to $NbF_5$ and $Si_2H_6$ at 200°C using the timing sequence 2-10-1-10. (b) Expanded view showing correlation between the QCM signal (solid line) and exposure to $NbF_5$ (dotted line) and $Si_2H_6$ (dashed line). The mass gain after one $Si_2H_6$ pulse is represented by $\Delta m_1$ and for a complete NbSi cycle by $\Delta m_2$.

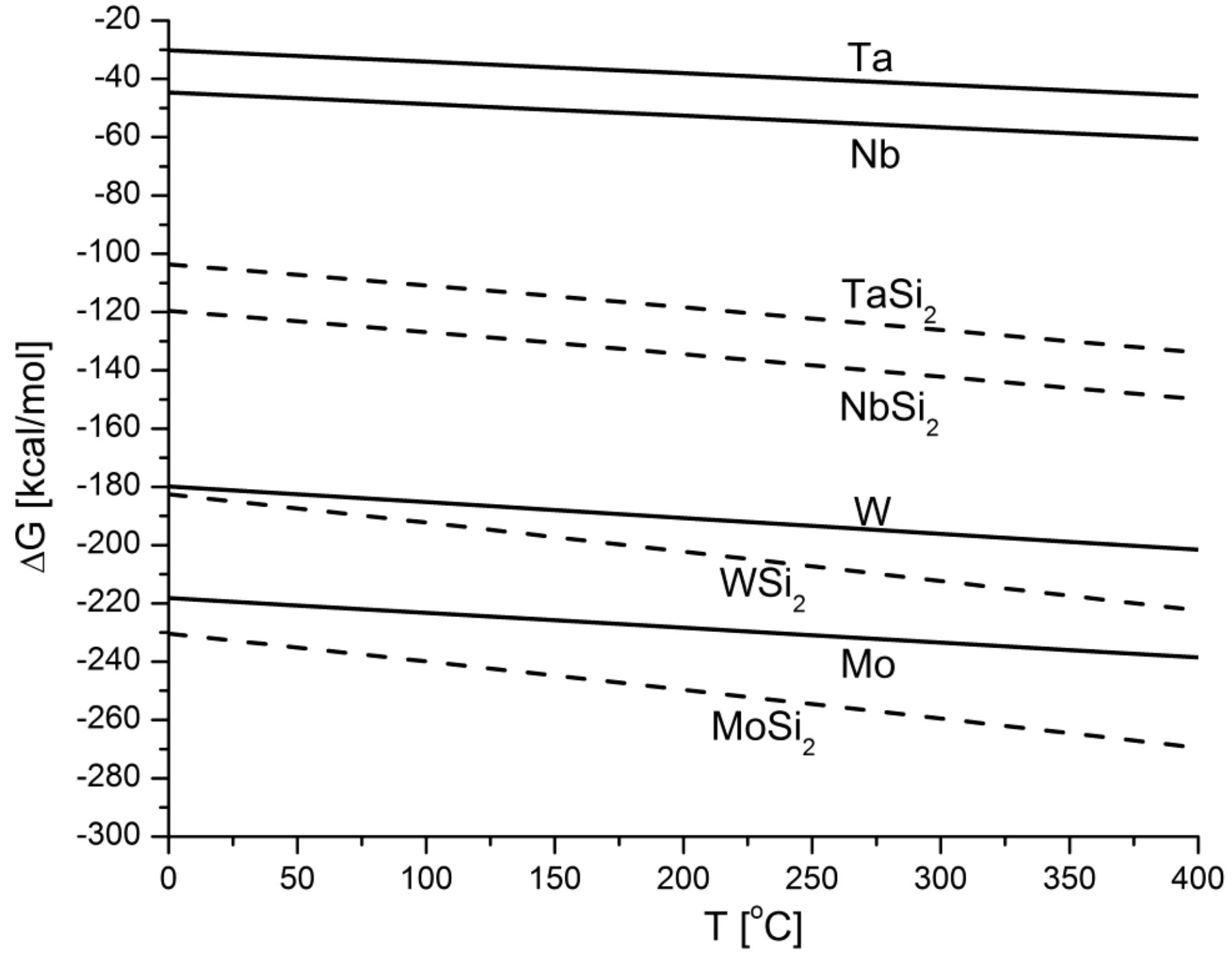

Figure 3: Enthalpy of formation per mole of reaction product as a function of the growth temperature for metal and metal silicide according to the following chemical reactions where the metals are represented by X:

Ta or Nb: $2XF_5 + Si_2H_6 = 2X + SiF_4 + SiHF_3 + H_2 + 3HF$

$TaSi_2$ or $NbSi_2$: $XF_5 + 3/2\ Si_2H_6 = XSi_2 + SiF_4 + 4H_2 + HF$

W or Mo: $2XF_6 + 3/2\ Si_2H_6 = 2X + 2SiF_4 + SiHF_3 + 7/2\ H_2 + HF$ [20]

$WSi_2$ or $MoSi_2$: $XF_6 + 3/2\ Si_2H_6 = XSi_2 + SiF_4 + 7/2\ H_2 + 2HF$

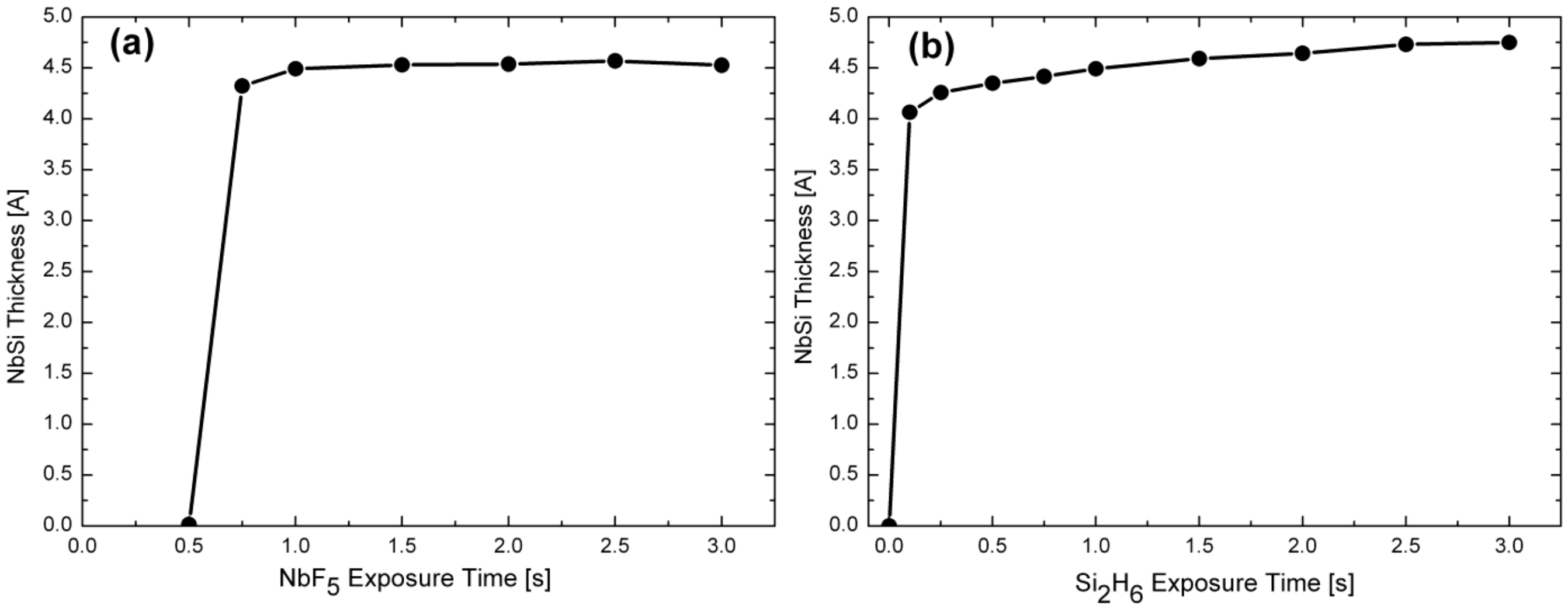

Figure 4: NbSi growth rates measured by XRR for films deposited on Si(100) at 200$^{o}$C (a) versus $NbF_5$ exposure time using the timing sequence x-10-1-10, and (b) versus $Si_2H_6$ exposure time using the timing sequence 2-10-x-10.

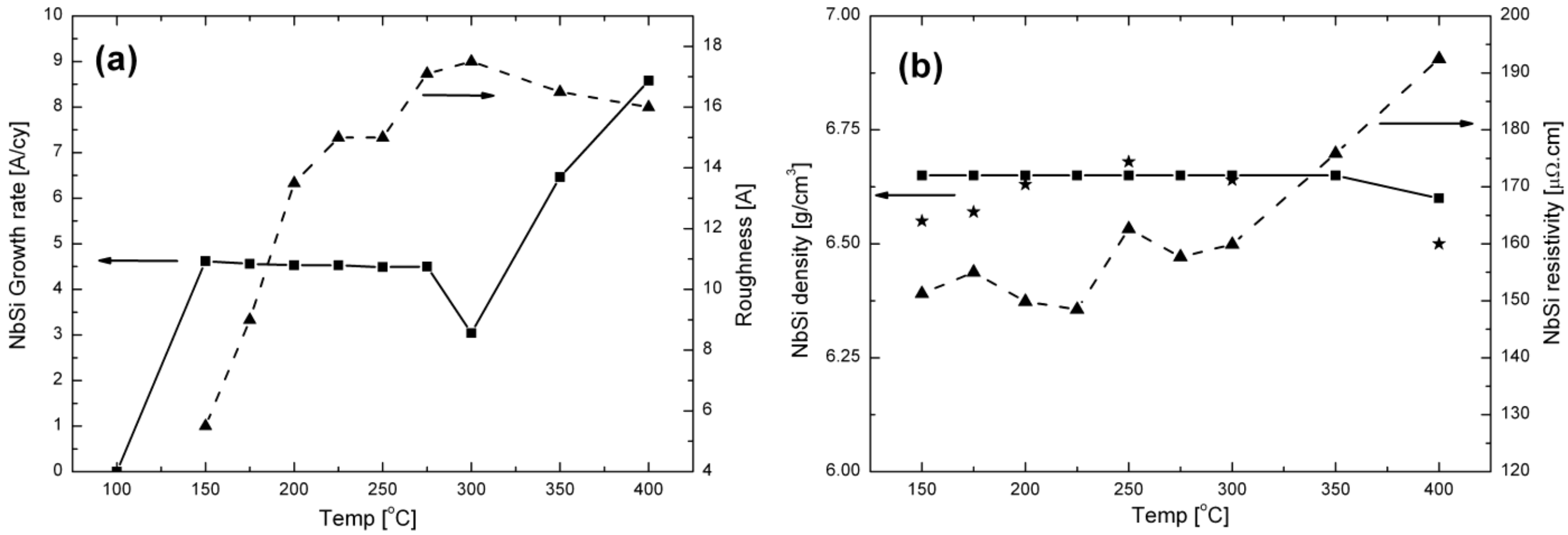

Figure 5: (a) NbSi growth rate and film roughness versus synthesis temperature measured by XRR for films deposited on Si(100) using 100 cy NbSi ALD with the timing sequence 2-10-2-10. (b) NbSi density measured by XRR (squares) and RBS (stars) and electrical resistivity versus the deposition temperature for the same films series.

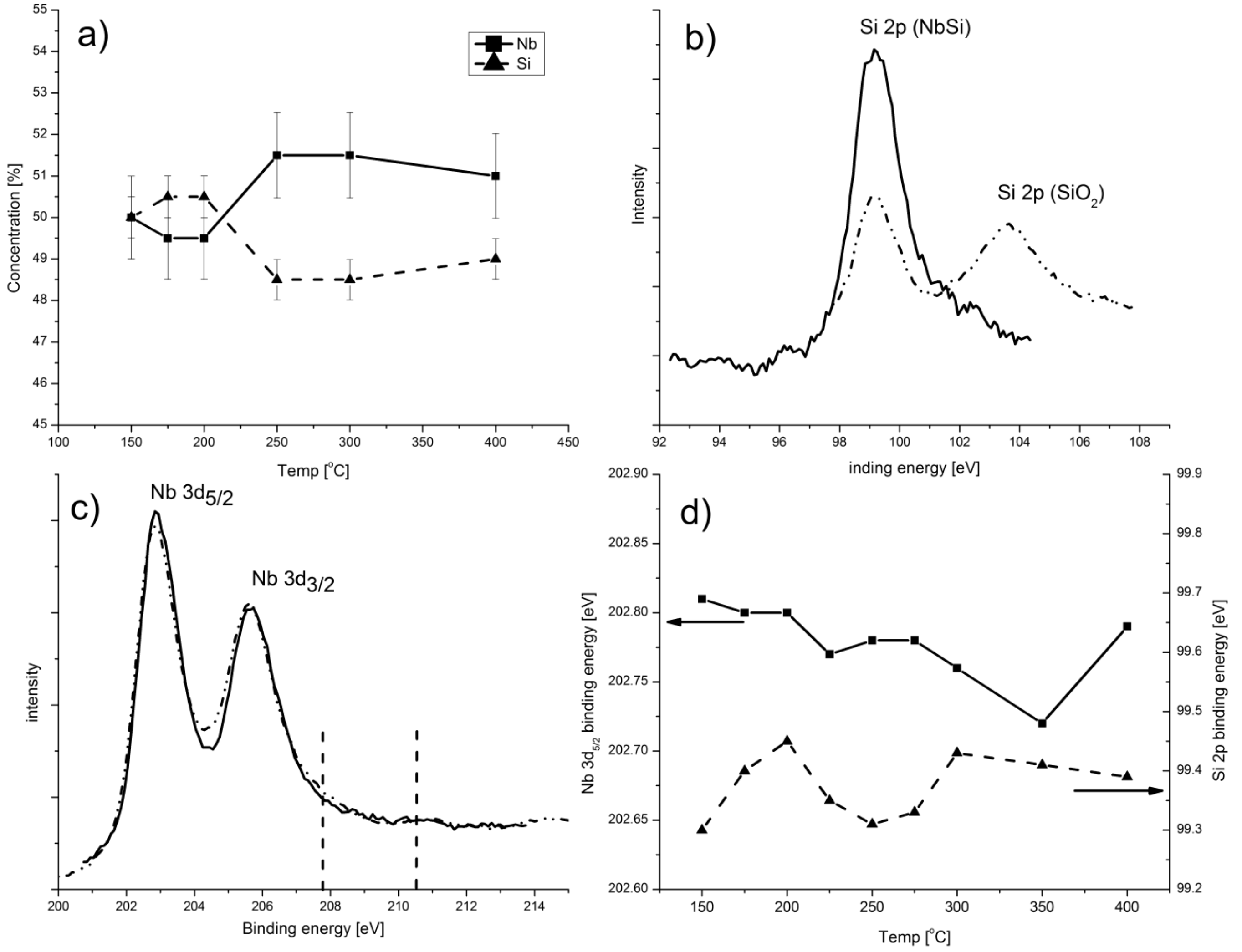

Figure 6: (a) Composition of NbSi films measured by RBS. (b) XPS spectra of Si 2p peak before (dash-dotted line) and after (solid line) Ar sputtering. (c) XPS spectra of the Nb 3d's peaks before (dash-dotted line) and after (solid line) Ar sputtering, the dotted lines indicate the Nb $3d_{5/2}$ and $3d_{3/2}$ XPS peaks positions for $Nb_2O_5$. (d) Binding energy of the Nb and Si peaks as a function of the growth temperature of films grown on Si(100) with 100Cy using the timing sequence 2-10-1-10.

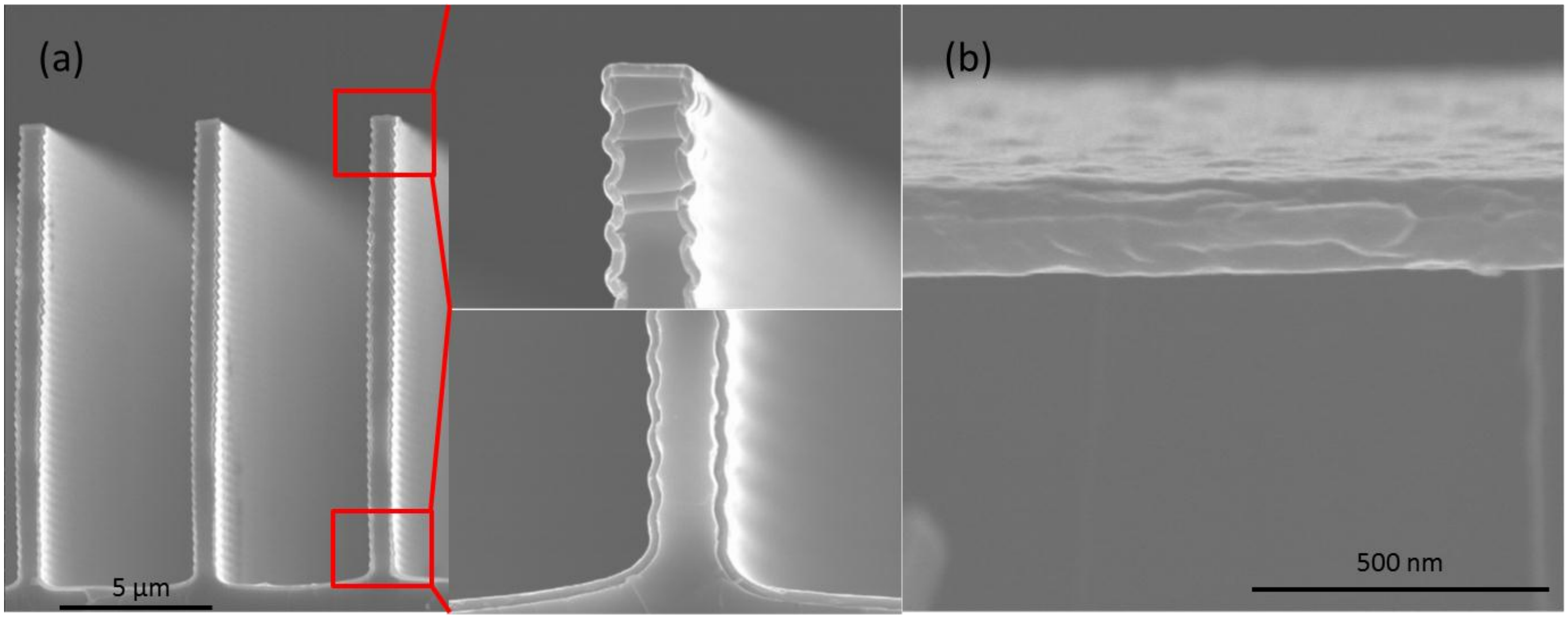

Figure 7: SEM pictures of a NbSi film grown on Si(100) at 200$^{o}$C using the pulsing sequence 2-10-1-10 and 300 ALD cycles. (a) on a trenched Si wafer (b) on a Si wafer.

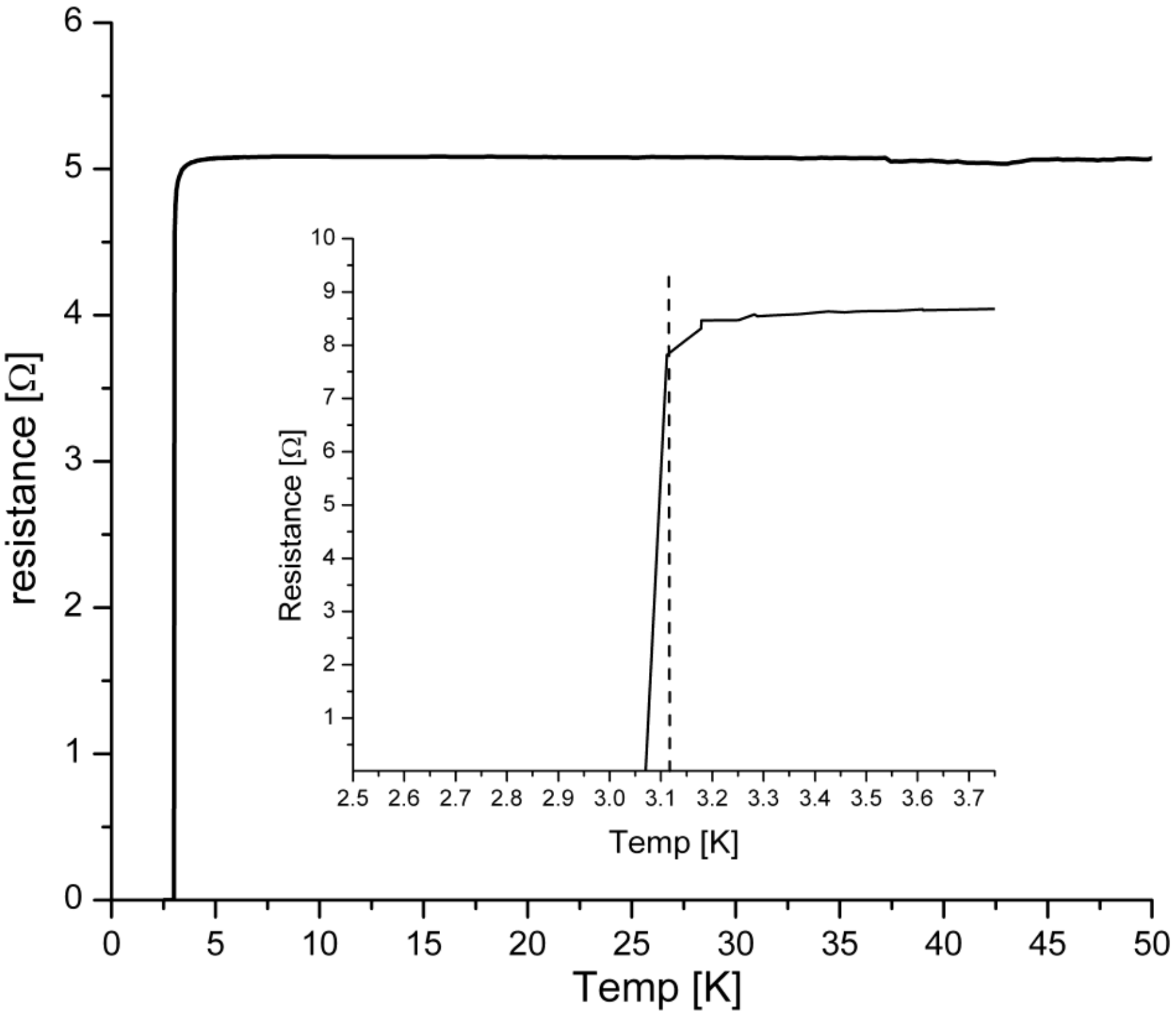

Figure 8: resistance versus temperature of a NbSi films grown at 200$^{o}$C on Si(100) using the timing sequence 2-10-1-10 and 100 ALD cycles. The insert is an expanded view near the superconducting transition, indicated by a dashed line.

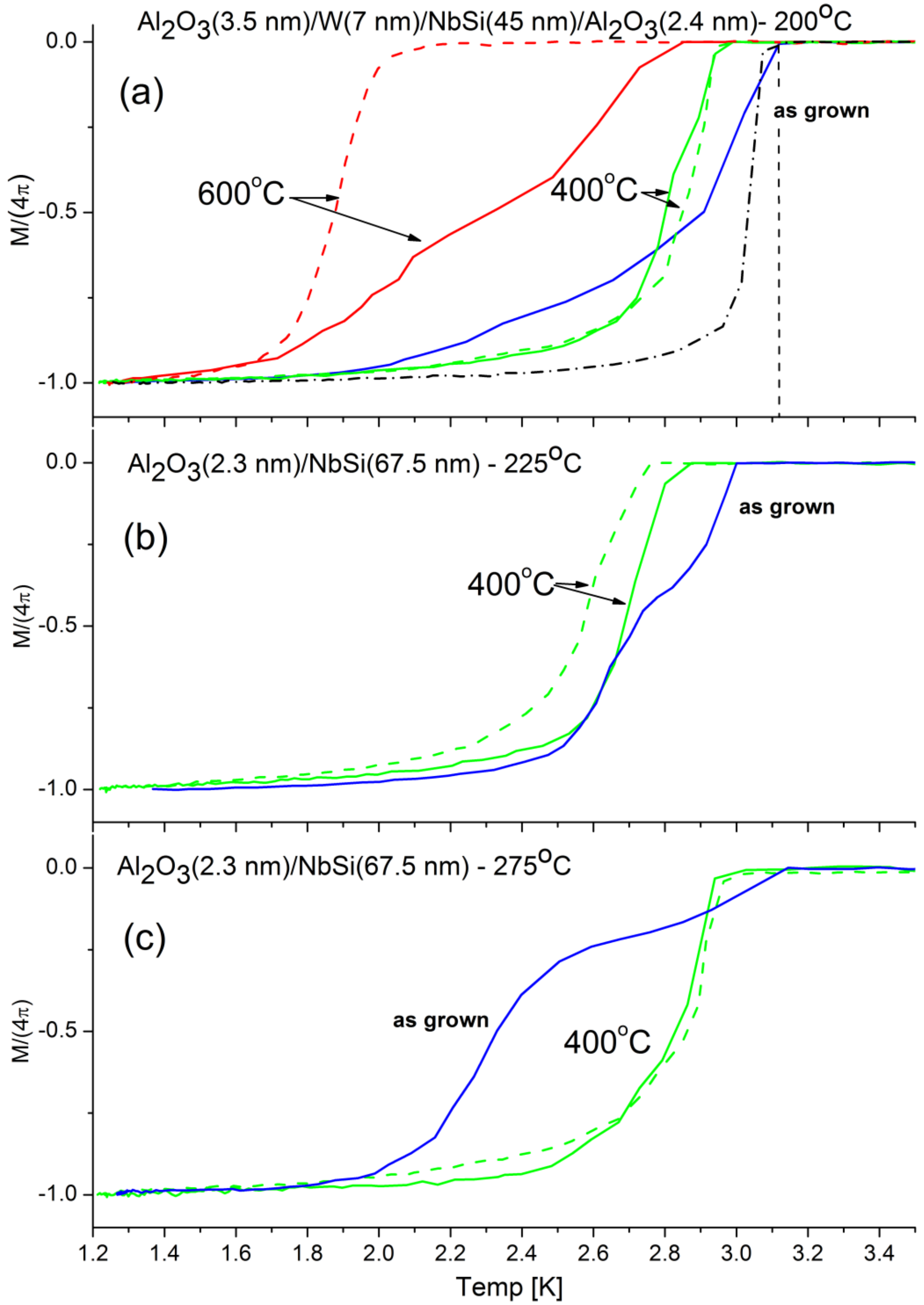

Figure 9: SQUID magnetometry measurements on NbSi films grown on Si(100) using the timing sequence 2-10-1-10. The curves labeled "as grown" correspond to measurements done on as grown films, the dashed lines correspond to measurement done after a post annealed films in Ar and the dotted lines in $N_2$. The post annealing temperatures: 400°C or 600°C are shown next to the corresponding curves.